\documentclass[11pt,a4paper]{article}
\usepackage[round]{natbib}

\usepackage{a4wide}
\usepackage{booktabs}
\usepackage[a4paper,top=2.5cm,bottom=2.5cm,left=3.5cm,right=2cm]{geometry}
\usepackage[utf8]{inputenc}
\usepackage[T1]{fontenc}
\usepackage{graphicx}
\usepackage{url}
\usepackage[hidelinks,breaklinks]{hyperref}
\usepackage{amsthm}
\usepackage{amsmath}
\usepackage{amssymb}
\usepackage{authblk}
\usepackage{float}
\usepackage{epstopdf}
\AppendGraphicsExtensions{.tif}

\usepackage{esvect}
\usepackage{subfig}
\usepackage[table,xcdraw]{xcolor}

\theoremstyle{definition}

\title{Innovative method for reducing uninformative calls in non-invasive prenatal testing}
\author[1,2,3]{Jaroslav Budis}
\author[2,4]{Juraj Gazdarica}
\author[2,5]{Jan Radvanszky}
\author[6]{Gabor Szucs}
\author[7]{Marcel Kucharik}
\author[2,4]{Lucia Strieskova}
\author[4]{Iveta Gazdaricova}
\author[2,4]{Maria Harsanyova}
\author[2,3]{Frantisek Duris}
\author[7]{Gabriel Minarik}
\author[7]{Martina Sekelska}
\author[8]{Balint Nagy}
\author[3,4,9]{Jan Turna}
\author[2,4,9]{Tomas Szemes}
\affil[1]{Department of Computer Science, Faculty of Mathematics, Physics and Informatics, Comenius University, Bratislava, Slovakia}
\affil[2]{Geneton s.r.o., Bratislava, Slovakia}
\affil[3]{Slovak Centre of Scientific and Technical Information, Bratislava, Slovakia}
\affil[4]{Department of Molecular Biology, Faculty of Natural Sciences, Comenius University, Bratislava, Slovakia}
\affil[5]{Institute of Clinical and Translational Research, Biomedical Research Center, Slovak Academy of Sciences Bratislava, Slovakia}
\affil[6]{Department of Applied Mathematics and Statistics, Faculty of Mathematics, Physics and Informatics, Comenius University, Bratislava, Slovakia}
\affil[7]{Medirex a.s., Bratislava, Slovakia}
\affil[8]{Department of Human Genetics, University of Debrecen, Debrecen, Hungary}
\affil[9]{Comenius University Science Park, Bratislava, Slovakia}
\date{}                     
\begin{document}

\maketitle
\begin{abstract}
{\bf Motivation.} Non-invasive prenatal testing or NIPT is currently among the top researched topic in obstetric care. While the performance of the current state-of-the-art NIPT solutions achieve high sensitivity and specificity, they still struggle with a considerable number of samples that cannot be concluded with certainty. Such uninformative results are often subject to repeated blood sampling and re-analysis, usually after two weeks, and this period may cause a stress to the future mothers as well as increase the overall cost of the test. 

{\bf Results.} We propose a supplementary method to traditional z-scores to reduce the number of such uninformative calls. The method is based on a novel analysis of the length profile of circulating cell free DNA which compares the change in such profiles when random-based and length-based elimination of some fragments is performed. The proposed method is not as accurate as the standard z-score; however, our results suggest that combination of these two independent methods correctly resolves a substantial portion of healthy samples with an uninformative result. Additionally, we discuss how the proposed method can be used to identify maternal aberrations, thus reducing the risk of false positive and false negative calls.

{\bf Availability and Implementation. } A particular implementation of the proposed methods is not provided with the manuscript.

{\bf Contact. } Correspondence regarding the manuscript should be directed at Frantisek Duris (fduris@dcs.fmph.uniba.sk).

{\bf Supplementary Information. } No additional supplementary information is available.
\end{abstract}



\section{Introduction}

Prenatal screening and diagnostics are important parts of obstetric care. Current methods of prenatal testing still involve most commonly invasive sampling of fetal material using procedures such as amniocentesis and chorionic villus sampling which are associated with a small but real risk of miscarriage $0.5 - 1\%$ \citep{mujezinovic2007procedure}. To prevent the risk of abortion associated with invasive sampling procedures, non-invasive prenatal testing (NIPT) based on fetal DNA analysis from maternal circulation has been developed. In 1997, the discovery of fetal cell-free DNA (cfDNA) in maternal plasma and serum revolutionized the area of non-invasive prenatal diagnostics, and opened up new options in the field of obstetric research \citep{lo1997presence}. The fetal cfDNA is of placental origin \citep{bischoff2005cell}, and it can be reliably detected from fifth week of gestation \citep{lo1998quantitative}. On average, the fetal cfDNA contributes about 10\% of all cfDNA fragments circulating in woman's blood when sampling is carried out between 10 and 20 gestational weeks, although the variance is quite large \citep{fiorentino2016importance}. The advance of massively parallel sequencing technologies together with the rapid development of bioinformatic algorithms and tools ushered in a new era of non-invasive prenatal identification of common fetal aneuploidies, now commonly known as NIPT \citep{chiu2008noninvasive,fan2008noninvasive,chiu2011non,sehnert2011optimal,bianchi2012genome,straver2013wisecondor,stephanie2014size,tynan2016application}.

While the performance of the current state-of-the-art NIPT solutions achieve high sensitivity and specificity \citep{bianchi2014dna,koumbaris2016cell}, they still struggle with a considerable number of samples that cannot be concluded with certainty. The great source of such uninformative samples is in the nature of the statistical testing. Considering standard cut-off threshold 2.5 for reliable conclusion of healthy samples \citep{bianchi2014dna}, and testing normally distributed ratios measured for the common aneuploidy chromosomes, the chance that a healthy sample would achieve z-score greater than this is around 1.86\%, and it is even higher when testing for other aberrations such as monosomy, gonosomal or sub-chromosomal aberrations.

Other problems are represented by maternal DNA aberrations such as maternal mosaicism \citep{wang2014maternal}, unidentified maternal tumors \citep{amant2015presymptomatic}, or copy number variations \citep{snyder2015copy,zhou2017contribution}. Hypothetically, a duplication of even a small part of maternal chromosome, which may not be detrimental for the mother, may result in a false positive call for fetal aneuploidy. This is because such duplication effectively increases the size of that chromosome, and, because maternal cfDNA is by far dominant, the signal from partial maternal duplication can be interpreted as full fetal trisomy. Similarly, an opposite effect can cause a false negative result.

There is a growing body of studies addressing this issue \citep{wang2014maternal,wang2015maternal,leilei2015maternal}. A particularly interesting venue of research focuses on qualitative differences between fetal and maternal cfDNA fragments, namely their lengths. It was previously reported that fetal fragments are on average shorter than maternal \citep{fan2010analysis,stephanie2014size,minarik2015utilization}. Using this information, \citet{shubina2017silico} were able to identify, post test, which of the false positive trisomy X samples were due maternal mosaicism. Briefly, they observed that in case of true fetal aneuploidy, the fetal fraction calculated from X chromosome increases when the long reads are filtered out. On the other hand, for maternal mosaicism the filtering has almost no effect on this fetal fraction.

Extending the above mentioned work, in this paper we present a method to further boost the elimination of uninformative results which have a potential to cause needless stress to the parents, requiring repeated blood samplings and analyzes. The associated increase of expenses, in turn, lower viability of the NIPT product. Reducing such cases is, therefore, of high interest in the area of NIPT. Our method is also based on the length of cfDNA fragments, taking into account how fragments of a particular length contribute to the z-score calculated from some chromosomes as in \citep{sehnert2011optimal}. Even if the proposed method is not as accurate as the standard z-score, our results suggest that combination of these two independent methods correctly resolves a substantial portion of healthy samples with an uninformative result. Additionally, we discuss application of the novel method for distinguishing between fetal (e.g., aneuploidy) and maternal signal (e.g., copy number variation).


\section{Material and methods}

\subsection{Sample acquisition}

We have collected altogether 2,621 samples with singleton pregnancy, of which 2,569 were negative for trisomy of chromosomes 13, 18 and 21, while 5 were confirmed as T13, 6 were confirmed as T18, and 39 were confirmed as T21. One negative sample (analyzed twice) was falsely reported as T18. The samples were predominantly of Slovak and Czech origin. All women participating in this study gave informed written consent consistent with the Helsinki declaration. 

\subsection{Sample preparation and sequencing}

Blood from pregnant women was collected into EDTA tubes and kept at $4^{\circ}C$ temperature until plasma separation. Blood plasma was separated within 36 hours after collection and stored at $-20^{\circ}C$ unit DNA isolation. DNA was isolated using Qiagen DNA Blood Mini kit. Standard fragment libraries for massively parallel sequencing were prepared from isolated DNA using an Illumina TruSeq Nano kit and a modified protocol described previously \citep{minarik2015utilization}. Briefly, to decrease laboratory costs, we used reduced volumes of reagents what was compensated by 9 cycles of PCR instead of 8 as per protocol. Physical size selection of cfDNA fragments was performed using specific volumes of magnetic beads in order to enrich fetal fraction. Illumina NextSeq 500/550 High Output Kit v2 (75 cycles) was used for massively parallel sequencing of prepared libraries using pair-end sequencing with read length of 2x35bp on an Illummina NextSeq 500 platform.

\subsection{Mapping and read count correction} \label{sec:mapping}

Sequencing reads were aligned to the human reference genome (hg19) using Bowtie 2 algorithm \citep{langmead2009ultrafast}. The first stage of data processing was carried out as previously described \citep{minarik2015utilization}. NextSeq-produced fastq files (two per sample; R1 and R2) were directly mapped using the Bowtie 2 algorithm with very-sensitive option. Unless stated otherwise, only randomly chosen 5 million of alignments for each sample were considered, thus reducing the between-sample variability induced by sequencing. Reads with mapping quality of 40 or higher were retained for further data processing. Next, for each sample the unique reads were processed to eliminate the GC bias according to (Liao et al., 2014) with the exclusion of intrarun normalization. Briefly, for each sample the number of unique reads from each 20kbp bin on each chromosome was counted. With empty bins filtered out, the locally weighted scatterplot smoothing (LOESS) regression was used to predict the expected read count for each bin based on its GC content. The LOESS-corrected read count for a particular bin was then calculated as $RC_{cor} = RC - |RC_{loess} - RC_{avg}|$, where $RC_{avg}$ is the global average of read counts through all bins, $RC_{loess}$ is the fitted read count of that bin, and $RC$ is its observed read count.

Furthermore, variability of human genome in population also contributes to the mapping bias, mainly in regions with common structural differences. At first, bin counts were transformed into a principal space, where the first component represented the highest variability across individuals in the control set. To normalize the sample, bin counts corresponding to predefined number of top components were removed to reduce common noise in euploid samples \citep{price2006principal,zhao2015detection,johansson2017novel}.

\subsection{Reference z-score calculation} \label{sec:traditional}

The reference z-scores of samples were calculated as normalized chromosome values (NCV) according to \citet{sehnert2011optimal}. Given our training set, the optimal reference chromosomes were determined to be 1, 4, 8, 10, 19 and 20 for trisomy 21, 4, 7, 8, 9, 10 and 16 for trisomy 18, and 3, 4 and 7 for trisomy 13. Similarly to \citep{bianchi2012genome}, samples scoring 4 and higher were considered trisomic, while samples scoring 2.5 or lower were considered euploid. The range (2.5, 4) was considered uninformative. We will refer to these NCV values as reference z-scores or $Z_{NCV}$.

\subsection{Length score calculation} \label{sec:length}

We defined three novel statistics based on the fragment lengths, each building on the previous one. The basis for our first novel statistic, termed $\lambda$-score, were read counts for the chromosome of interest (e.g., the usual $13^{th}$, $18^{th}$ or $21^{st}$ chromosome). However, in contrast with the traditional z-scores, we did not compare the read count with the expected normal value estimated from a set of euploid samples. Instead, we continuously eliminated fragments of certain length and compared the observed counts with the expected counts, if the same amount of fragments was eliminated randomly.

More particularly, if $i$ marks the chromosome of interest and $chr_i$ the number of reads mapped to the $i^{th}$ chromosome of the tested sample, then let $chr_i(l)$ be the number of reads mapped to the $i^{th}$ chromosome that are of length at most $l$. Furthermore, let the total number of reads of the sample be $n$, and let $n_l$ be the total number of reads that are of length at most $l$. The number of reads mapped to the $i^{th}$ chromosome (of any length) after uniform random elimination follows a binomial distribution with parameters $(n_l,p)$, $p=chr_i/n$. The ratio $chr_i/n$ gives the proportion of reads originally mapped to the $i^{th}$ chromosome, and the number $n_l$ gives the number of reads to be drawn. Thus, the expected number of reads mapped the $i^{th}$ chromosome after the uniform random draw and its variance is equal to $e_l = n_l p$ and $v_l = n_l p (1 - p)$ as in binomial distribution, respectively. 

For each sample and chromosome, we defined a series of $\lambda$-scores as
\begin{equation}
\lambda_i(l) =  \frac{chr_i(l) - n_lp}{\sqrt{n_lp(1-p)}},
\end{equation}
where $l$ ranges from 50 to 200 and $i$ indicates a chromosome. Next, we defined a second novel statistic termed $FL$-score as
\begin{equation}
FL_i = \max_{125 \leq l \leq 145} \lambda_i(l).
\end{equation}
The bounds 125 and 145 were determined empirically. Finally, the $FL$-scores were normalized into z-scores which approximately follow standard normal distribution for euploid samples. Then, for any sample, its normalized $FL$-score value $Z_{FL}$ was used as an alternative method for the prediction of aneuploidy.

\subsection{Combining the scores} \label{sec:combi}

In our analysis, we found no correlation between the reference ($Z_{NCV}$) and length-based ($Z_{FL}$) scores measured in our training set of euploid samples $(\text{Pearson } R=0.017, p = 0.129)$. Thus, we considered them as two independent random variables, each from a standard normal distribution. The sum of their squares follows a chi-squared distribution with 2 degrees of freedom, and a survival function of the chi-squared distribution was used to associate this sum with the probability. Finally, this probability was converted back to a standard score $Z_{NCV+FL}$ through quantile function for easy comparison with other methods. Note that the calculations were performed in log-space to overcome underflow issues.

\begin{figure}
    \centering
    \subfloat[Combination of statistically independent NCV and FL scores. The scores were combined through $\chi^2$ distribution with 2 degrees of freedom (section \ref{sec:combi}).]{{\includegraphics[width=.6\linewidth]{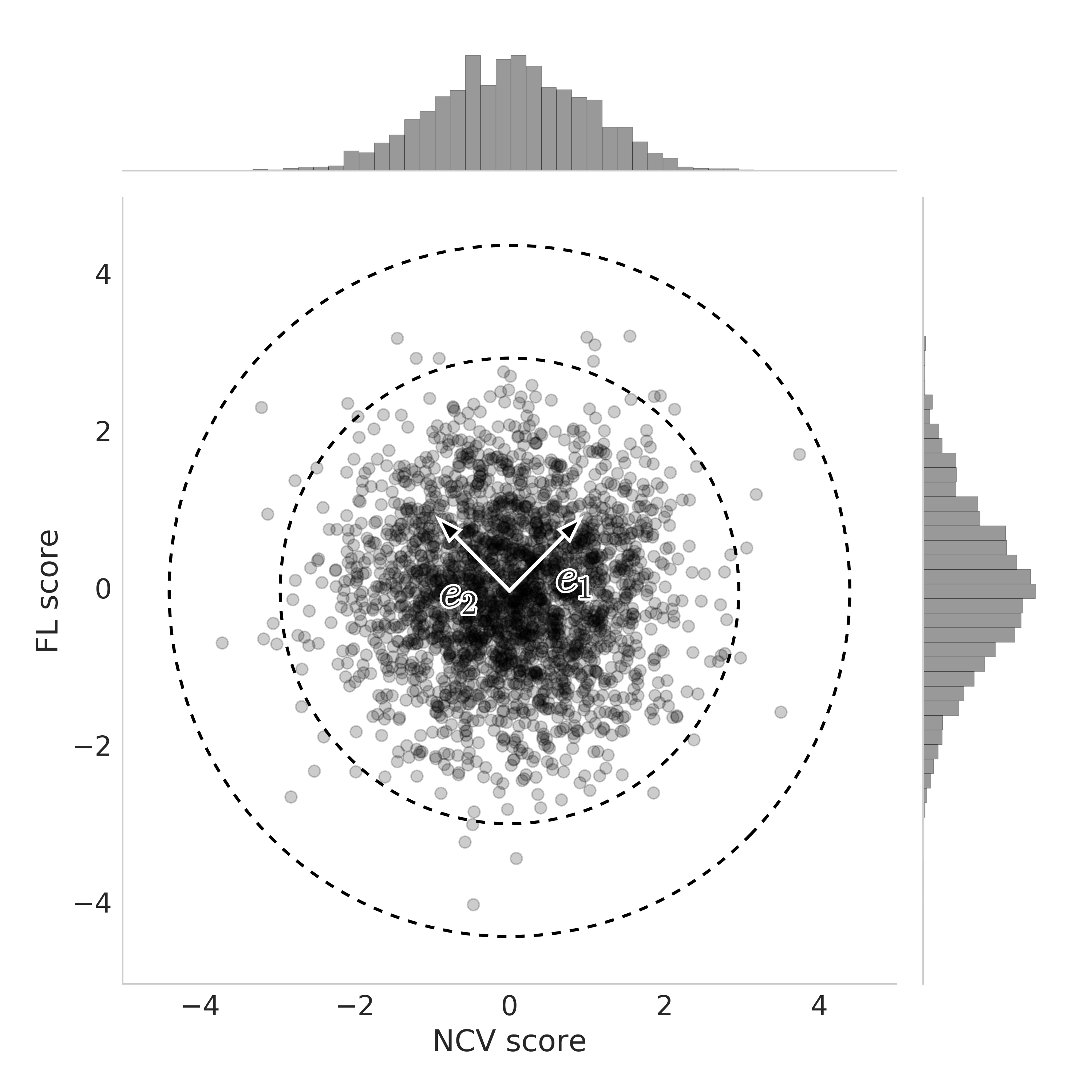} }}
    \qquad
    \subfloat[Combination of statistically dependent SZ and FL scores. First, principal component analysis was used to calculate axes $(e_1,e_2)$ of variation along which the two principal components were independent. Subsequently, $\chi^2$ distribution with 2 degrees of freedom was used to combine the principal components (section \ref{sec:combi}).]{{\includegraphics[width=.6\linewidth]{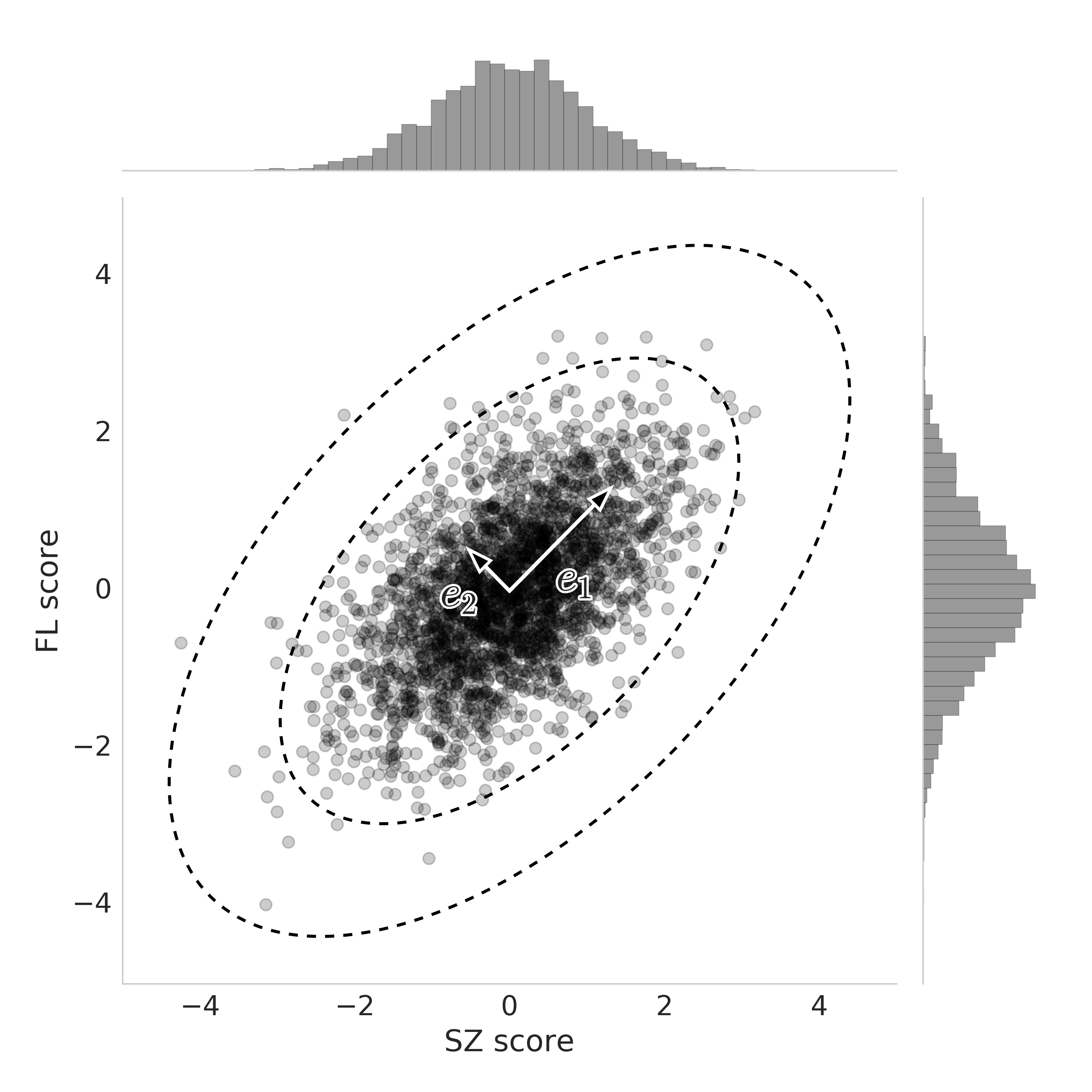} }}
    \caption{Combination of novel FL score with the reference score NCV (1a) and length-reduced reference score SZ (1b). Displayed are negative samples for chromosome 21. The ellipses represent combined z-scores 2.5 and 4, respectively.}
    \label{fig:combination}
\end{figure}

On the other hand, there was a significant correlation between $Z_{SZ}$ (defined below in section \ref{sec:silico}) and $Z_{FL}$ $(\text{Pearson } R=0.55, p < 0.001)$. In this case, we combined them according to \citep{owen1985canonical}. Specifically, we first performed a principal component analysis of the pairs $\left(Z_{SZ}, Z_{FL}\right)$. Then, we recalculated the z-scores along the newly found eigenvectors $e_1,e_2$ using the respective eigenvalues as new variances. The resulting scores were two independent standard normals, and we proceeded to combine them as in the first paragraph of this section. The resulting z-score was marked as $Z_{SZ+FL}$ (Figure \ref{fig:combination}).

\subsection{Statistical analysis}

The significance of our findings was evaluated using statistical tests implemented in Python \emph{scipy} package \citep{jones2014scipy}. The linear dependency of the two scores in negative samples was calculated with Pearson correlation. Since scores of aberrant samples were not normally distributed, Wilcoxon signed-rank test was used to estimate statistical significance of improvement between the reference and proposed methods.

\section{Results}

\subsection{$\lambda-$score profiles}

\begin{figure}
\centerline{\includegraphics[width=\textwidth]{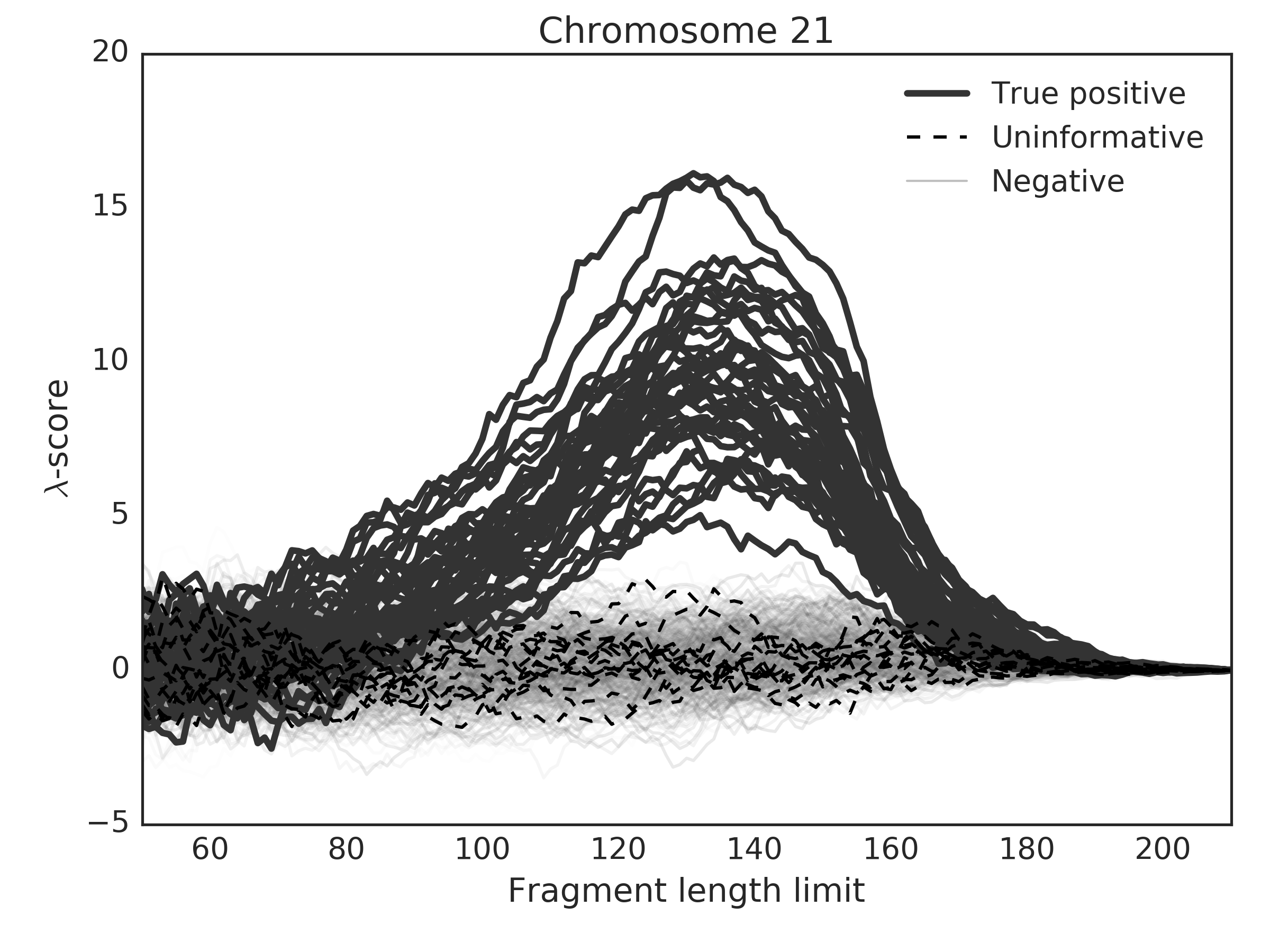}}
\caption{Deviation of the T21 positive samples from the euploid samples in terms of $\lambda$-score. Elimination of fragments with length above a threshold (Fragment length limit) leads to divergence of chromosomal counts from the expectations. In the healthy samples (only 500 samples shown for clarity), the difference can be explained by fluctuations in random draw. On the other hand, aberrant samples show profile that markedly diverge from expected values.}
\label{fig:lambda}
\end{figure}

First, we calculated series of $\lambda$-scores for euploid and trisomic samples for chromosomes 13, 18 and 21, and length range from 50bp to 220bp. We observed that trisomic samples behaved differently than euploid samples (Figure \ref{fig:lambda}). This difference can be explained by fetal fragments being shorter than maternal. The fetal and maternal cfDNA differ in properties such as length distribution and source chromosomes \citep{fan2010analysis,stephanie2014size}. Originally, they are mixed in some ratio, usually termed fetal fraction, resulting in the observed properties of the mixture (read length distribution and chromosome mapping ratios). Uniform random elimination of fragments preserves these properties of the mixture on account of being uniform, both in euploid and trisomic cases. On the other hand, elimination by length as described in section \ref{sec:length} eliminates more maternal than fetal fragments on account of the latter being shorter than the former. Thus, the original properties of the mixture are not preserved, if fetal and maternal properties are not the same, which is the case for trisomic samples with respect to chromosome mapping ratios.

At first, the $\lambda$-scores of trisomic samples gradually increase with the elimination of longer fragments. This positive effect is however balanced by the negative effect of lower number of remaining fragments, and so, after while, the $\lambda$-scores decline to the values expected by a random draw. We observed the highest deviation of aberrant samples using thresholds for fragment lengths between 125bp and 145bp. We therefore measured the maximal value in this range which we termed FL score of the sample.

Note that in contrast with the reference z-score $Z_{NCV}$, the calculation of $\lambda$-scores is not based on comparison of proportions of fragments with healthy population. An excessive number of fragments from a chromosome would therefore not result in the positive call, if these fragments do not have fetal length distribution. Thus, it is possible to reveal false positive calls caused by maternal aberrations. Similar concepts have been already utilized in distinction between maternal and fetal gonosomal aberrations \citep{shubina2017silico}, leading to the reduction of false positive results of the monosomy X0 predictions (section \ref{sec:ivf}).

\subsection{\emph{In silico} size selection} \label{sec:silico}

Next, we examined the effect of the length-based fragment filtering on the reference z-scores. Particularly, we first calculated the reference z-scores $Z_{NCV}$ (section \ref{sec:traditional}), and then we removed the fragments longer than 150bp. However, this considerably changed the read count for our samples ($1.5 \pm 0.3$ million), so we concluded that the trained mean and standard deviation (used in z-score calculation) for the original data may not be suitable for the length-reduced data. Therefore, we applied this length-reduction to our training samples as well, determined the new mean and standard deviation (we kept the reference chromosomes the same), and only with these new values we calculated z-scores of the length-reduced test samples. Comparing the original and length-reduced z-scores (Figure \ref{fig:trisomy}), we observed only small and statistically insignificant increase in z-scores of trisomic samples (average multiplicative increase $1.05\times$; Wilcoxon $Z = 516, p < 0.241$). For future reference, we termed these z-scores $Z_{SZ}$.

On the other hand, when all aligned reads were considered (recall that up until now, all samples were restricted to the first 5 million raw alignments, see section \ref{sec:mapping}), and the same procedure was applied again, the increase in z-scores of trisomic samples became statistically significant (average multiplicative increase $1.13\times$; Wilcoxon $Z = 136, p < 0.001$). As before, we recalculated the mean and standard deviation used in the z-score calculation because of the changed read count per sample ($8.8 \pm 4.6$ million), while the reference chromosomes were again kept the same (section \ref{sec:traditional}).

This finding indicates that the length-based fragment selection is beneficial only for samples with more than 5 million aligned fragments, at least for the reference z-scores. This is in accord with our previous findings that \emph{in silico} size-based filtering of fragments (only reads up to 155bp were retained) did not lead to statistically significant increase in trisomic z-scores using MiSeq runs having $3.1 \pm 1.0$ million reads per sample, while statistically significant ($p = 0.04$) increase was observed for samples sequenced on Ion Torrent PGM, of which samples had $5.8 \pm 1.0$ million reads \citep{minarik2015utilization}.

\subsection{Combined scores as supplemental evaluation method}

\begin{figure}
\centerline{\includegraphics[width=\textwidth]{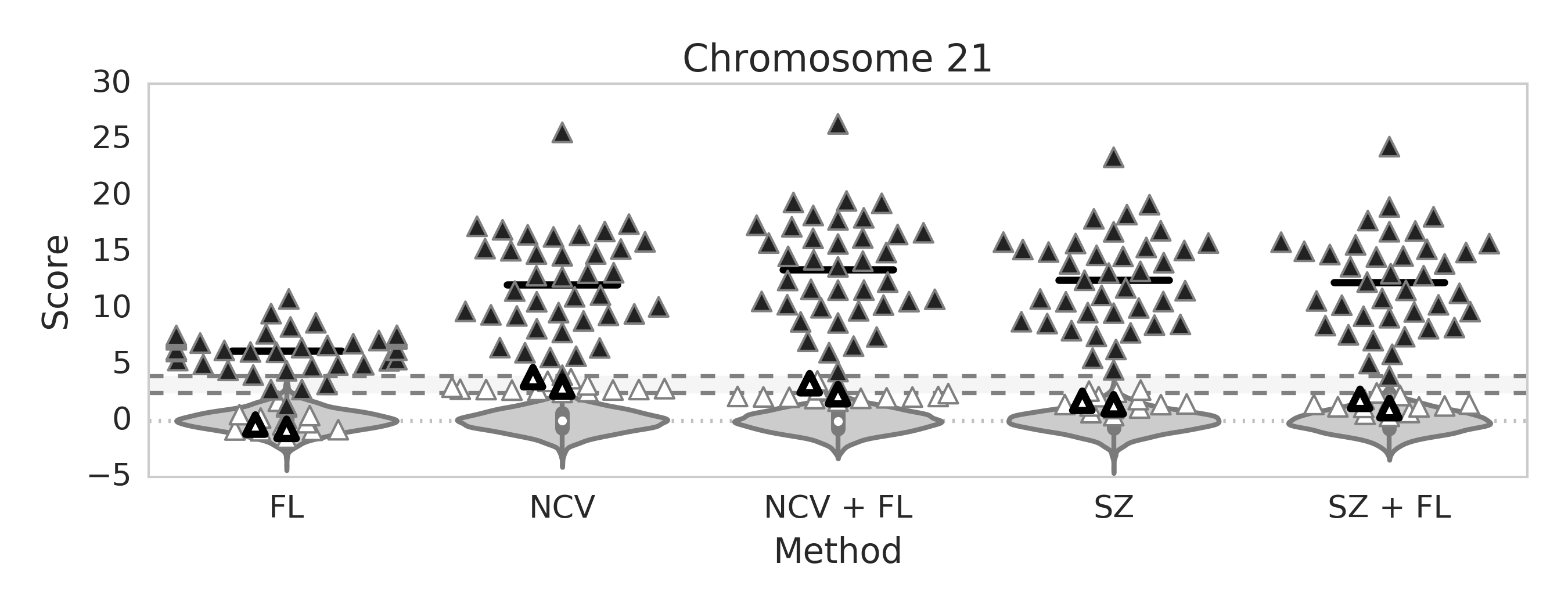}}
\centerline{\includegraphics[width=\textwidth]{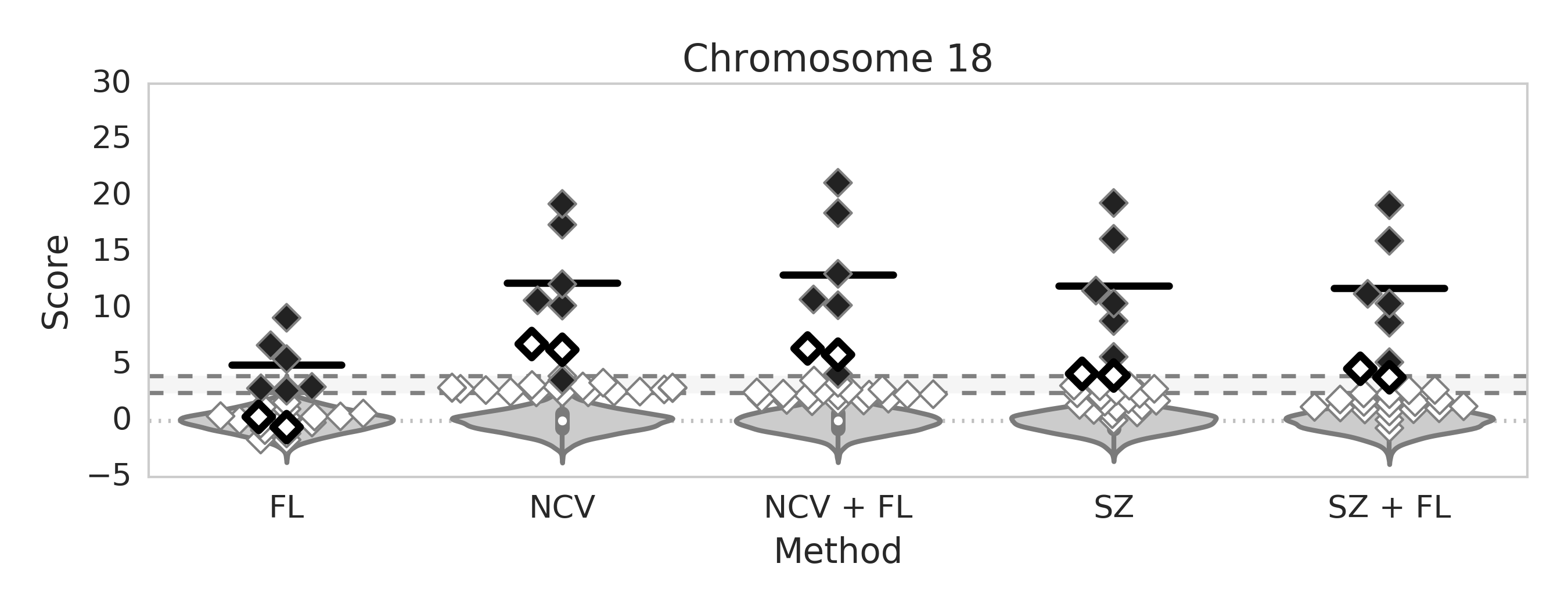}}
\centerline{\includegraphics[width=\textwidth]{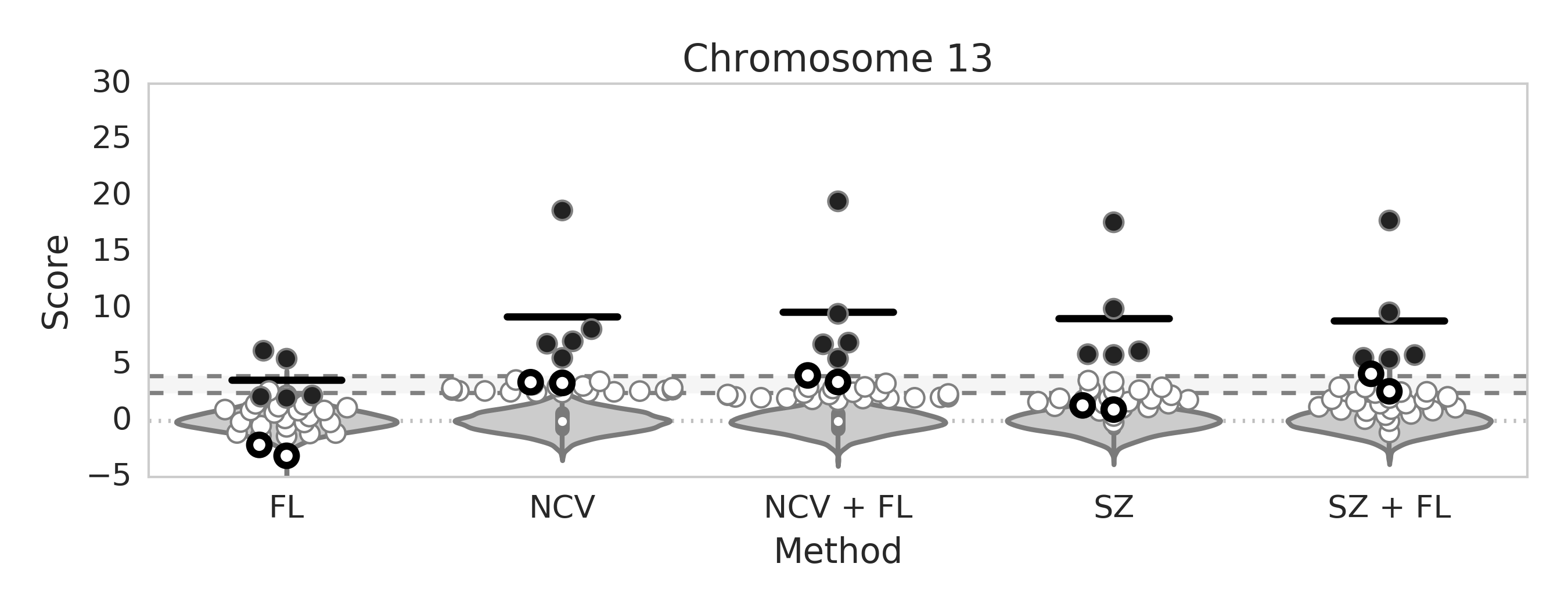}}
\caption{Comparison of z-score calculation methods on chromosome 21 (upper), chromosome 18 (middle) and chromosome 13 (lower). Reference method based on chromosomal fragment counts (NCV) performs better than proposed FL method, albeit their combination (NCV + FL) increased z-scores of trisomic samples (black marks). Size selection (SZ) further improves z-scores of trisomic samples. The combination with FL method (SZ + FL) most markedly reduces z-scores of samples evaluated as uninformative using traditional NCV score alone (white marks). Empty shapes represent results of the two analyzes of the IVF sample discussed in section \ref{sec:ivf}.}
\label{fig:trisomy}
\end{figure}

The performance of the novel statistic $Z_{FL}$ by itself was observed to be weaker than that of the traditional $Z_{NCV}$ (Figure \ref{fig:trisomy}). Particularly, there was a substantial decrease in z-scores for trisomic samples. On the other hand, their combination $Z_{NCV+FL}$ resulted in statistically significantly higher trisomic z-scores than $Z_{NCV}$ (average multiplicative increase $1.11\times$; Wilcoxon $Z = 17, p < 0.001$). Surprisingly, the combined score $Z_{SZ+FL}$ produced only statistically insignificant increase of trisomic z-scores (average multiplicative increase $1.02\times$; Wilcoxon $Z = 637, p > 0.05$).

Furthermore, we observed a decrease of $Z_{SZ+FL}$ scores of false positive and uninformative samples when compared with reference $Z_{NCV}$, even though other samples that were previously classified as negative replaced them in the uninformative range. Additionally, $Z_{NCV+FL}$ and $Z_{SZ+FL}$ resulted both in false positive results indicating that using these metrics alone may decrease accuracy of the testing.

\subsection{Improved evaluation method} \label{sec:method}

In light of our findings, we propose a following improvement of the NIPT evaluation process. First, a reference $Z_{NCV}$ is calculated. If it is in the negative zone, then no further test is applied and the sample is closed as normal. If the $Z_{NCV}$ is in the uninformative zone, then $Z_{SZ+FL}$ is computed, and if this new score is in the negative zone, the sample is closed as normal. Our findings indicate that this process considerably decrease the number of uninformative results (Figure \ref{fig:trisomy}, the white marks in particular).

\subsection{A false positive sample} \label{sec:ivf}

\begin{figure}
\centerline{\includegraphics[width=\textwidth]{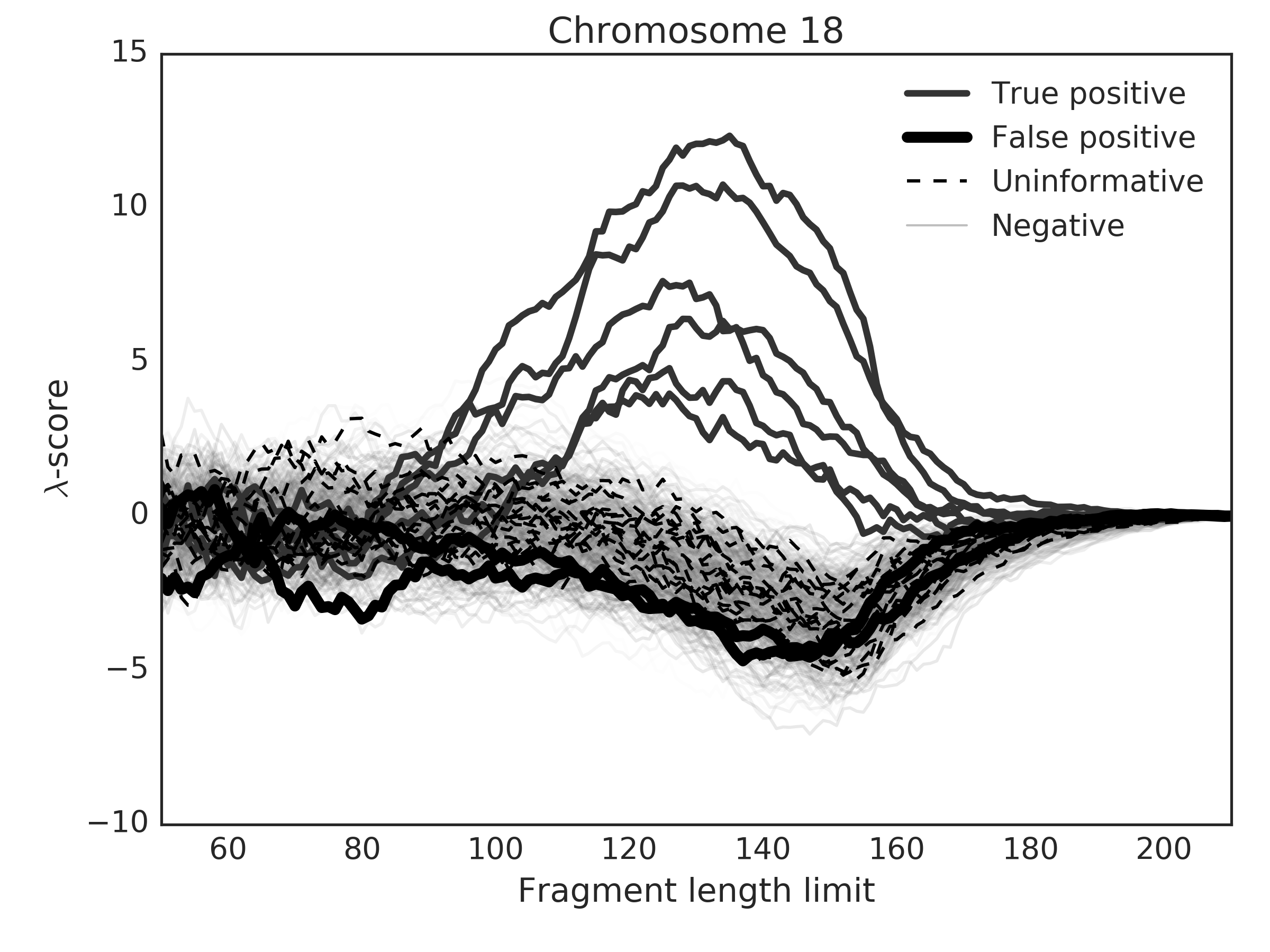}}
\caption{$\lambda$-scores for chromosome 18. Interesting is the twice analyzed false positive T18 sample discussed in section \ref{sec:ivf}. The $\lambda$ profile, similar to healthy samples, may indicate that the revealed aberration by $Z_{NCV}$ (6.82 and 6.31) may be mistaken.}
\label{fig:t18}
\end{figure}

We observed a atypical sample from IVF pregnancy (analyzed twice) with high risk for trisomy 18 (Figure \ref{fig:t18}). Although both analyzes were supported by solid $Z_{NCV}$ scores 6.82 and 6.31, the predicted aberration was not confirmed by invasive follow-up test. On the other hand, the fragment length $Z_{FL}$ score classified it as healthy sample (-0.57, 0.34). The combination of these scores $Z_{chi}$ led to a slightly reduced scores (6.43, 5.88), and the inclusion of the \emph{in silico} size selection further reduced their z-scores (3.64, 3.44), getting them into uninformative range.

The most likely explanation is that this was a result of a maternal copy number variation on the $18^{th}$ chromosome which is known to cause false positive results \citep{zhou2017contribution}. Unfortunately, we were not able to perform a follow-up of maternal genotype. If this was the case, it would indicate that the proposed method, particularly the $\lambda$ profiles, can be used to distinguish between fetal (e.g., aneuploidy) and maternal (e.g., copy number variation) NIPT signals. Such method would be very valuable, but validation on a larger cohort of such samples is required.

Other possible and rather interesting hypothesis is that this was the case of a trisomic vanishing twin. In this case, we would expect that the cfDNA fragment
length distribution originating from the trisomic vanishing twin would be similar to that of the trisomic living fetus, though we were not able to find publications supporting or opposing this expectation. However, the observation indicates that the fragment length distribution is similar to that of a mother. Thus, either the vanishing twin hypothesis or the assumption about the read length distribution is wrong. In case of the latter, such difference could then be utilized for reducing the false NIPT results due to vanishing twin effects.

\section{Conclusion}

The reference z-score method $(Z_{NCV})$ of aneuploidy prediction based on chromosomal proportions is already widely accepted, since it can quite well separate between trisomic and euploid samples. To avoid false predictions in routine diagnosis, NIPT tests have typically a range of scores which are considered too risky for definite predictions. Such uninformative results are often subject to repeated blood sampling and re-analysis, usually after two weeks, and this period may cause a stress to the future mothers. In addition to that, high numbers of uninformative results also increase the overall cost of the whole procedure. The supplementary scores proposed in this paper may offer a useful way for reducing the number of uninformative samples in several ways.

In our data set, we classified 53 of 2,569 (2.06\%) negative samples as uninformative using the reference $Z_{NCV}$ method. The combination of $Z_{NCV}$ with the $Z_{FL}$ score, $Z_{NCV+FL}$, led to similar number 51 (1.98\%), albeit only 15 of them were shared in both metrics. The addition of size selection to the combined method $(Z_{SZ+FL})$ increased number of uninformative samples to 48 (1.87\%) but only 4 of them were shared with the reference method. Thus, the combination of these approaches can be used to substantially reduce the number of uninformative results as proposed in section \ref{sec:method}. In our case, 49 out of 53 uninformative samples would be closed as negative without affecting the prediction of true positive samples.

Another way to lower the number of uninformative samples is to classify uninformative samples with low $Z_{FL}$ score as negative. Setting the threshold for $Z_{FL}$ score to 0, 1, 2 led to correct elimination of 24 (45.28\%), 41 (77.36\%) and 50 (94.34\%) out of 55 true negative samples with uninformative $Z_{NCV}$ call. The thresholds, however, must be chosen with caution, since the lowest observed $Z_{FL}$ score of positive sample was 1.34. Because this sample also had high chromosomal $Z_{NCV}$ score 11.02, the false negative call with $Z_{FL}$ threshold set to 2 could be avoided by preferring the reference method, albeit similarly low $Z_{FL}$ scores may occur along with the low score of the reference method.

\begin{figure}
\centerline{\includegraphics[width=\textwidth]{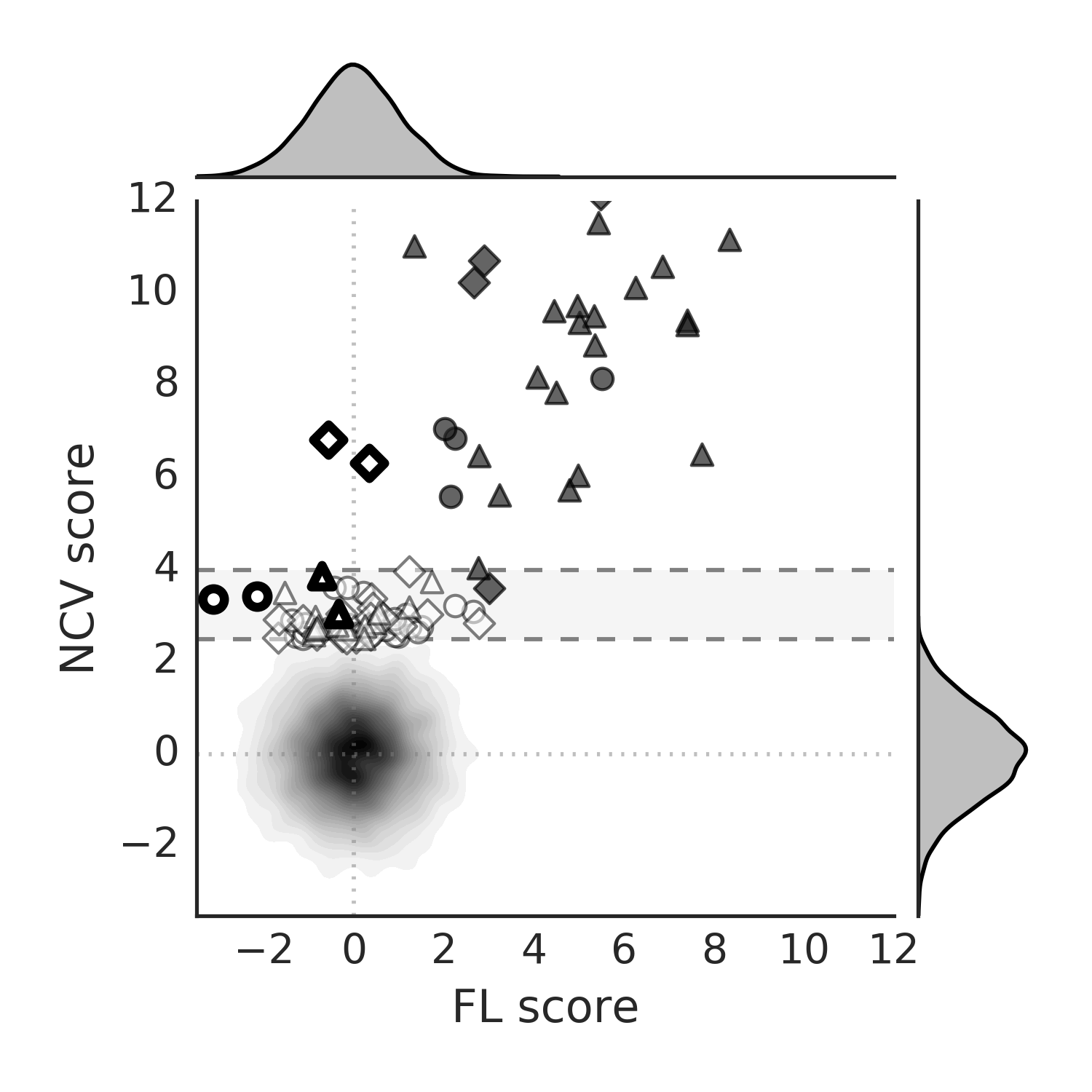}}
\caption{Comparison of z-scores calculated chromosomal count (NCV) and fragment length (FL) methods. Positive (black) and negative (white) samples with z-scores higher that 2.5 (NCV method) are represented by triangles (chromosome 21), diamonds (chromosome 18) and circles (chromosome 13). Other negative samples are accumulated in form of a density plot in the intersection of major axes ($x=0$; $y=0$). Major axes divide predictions into four classes, while the fetal trisomies are located in top, right quartile ($Z_{FL} > 0$, $Z_{NCV} > 0$). Empty shapes represent results of the two analyzes of the IVF sample discussed in section \ref{sec:ivf}.}
\label{fig:blot}
\end{figure}

A typical NIPT analysis is suitable for prediction of monosomy as well. The monosomic samples are distinguished by negative z-scores below some predefined threshold, for example $-3$ \citep{mazloom2013noninvasive}. In contrast with the sign of chromosomal $Z_{NCV}$ score, which indicate increase or decrease of DNA material from specific chromosome, the sign of the fragment length $Z_{FL}$ score indicates maternal or fetal origin of the aberration. This way, the scores may be divided into four categories representing: 1) maternal duplication ($Z_{FL} < 0$, $Z_{NCV} > 0$), 2) fetal duplication ($Z_{FL} > 0$, $Z_{NCV} > 0$), 3) maternal deletion ($Z_{FL} < 0$, $Z_{NCV} < 0$), and 4) fetal deletion ($Z_{FL} > 0$, $Z_{NCV} < 0$), thus making the prediction more informative (see Figure \ref{fig:blot}).

Also, we observed a statistically significant increase in trisomic z-scores for method $Z_{NCV+FL}$. Thus, this method can be considered as an improved version of $Z_{NCV}$.

Finally, we discuss the possibility of either our method being able to distinguish between fetal and maternal NIPT signals or trisomic vanishing twin having different cfDNA fragment length distribution than living trisomic fetus. To verify these hypotheses, more tests on a larger cohort of samples are required.

Elimination of long fragments may significantly improve prediction accuracy of trisomy testing. The number of sequenced DNA fragments must be however sufficient to balance the lower number of analyzed fragments. Based on the patterns observed in z-score profiles, we designed a novel method for prediction that is independent to standard method based on chromosomal counts. We presented that combination of these two methods may conclude samples that cannot be safely classified using a single method.

\section{Acknowledgments}

This contribution is the result of implementation of the project \emph{REVOGENE} $-$ \emph{Research centre for molecular genetics} (ITMS 26240220067) supported by the Research \& Developmental Operational Programme funded by the European Regional Development Fund.

\bibliographystyle{abbrvnat}
\bibliography{bibi}

\eject

\eject

\end{document}